\documentclass[11pt,a4paper]{article}

\usepackage[T1]{fontenc}
\usepackage[utf8]{inputenc}
\DeclareUnicodeCharacter{2212}{\ensuremath{-}}
\usepackage[version=4]{mhchem}
\usepackage{siunitx}
\usepackage{textgreek}
\usepackage{authblk}
\usepackage{amsmath,amssymb,amsthm}

\usepackage{graphicx}
\usepackage{float}

\usepackage[colorlinks=true,
            linkcolor=blue,
            citecolor=blue,
            urlcolor=blue]{hyperref}
\usepackage{cleveref}

\usepackage[backend=biber,
            style=numeric,
            doi=false,
            eprint=false,
            url=false,
            sorting=none
            ]{biblatex}
\usepackage[a4paper,
            margin=2.5cm]{geometry}

\usepackage{booktabs}

\title{Plastic Relaxation without Dislocations in $\beta$-\ce{Ga2O3} Heteroepitaxy: A Structural Peculiarity of \ce{Ga2O3} Polymorphs}

\author[1]{I. Bertoni\thanks{Corresponding author: i.bertoni1@campus.unimib.it}}
\author[1]{A. Marzegalli}
\author[1]{A. Ugolotti}
\author[2]{R. He}
\author[2]{F. Djurabekova}
\author[1]{L. Miglio}

\affil[1]{Department of Materials Science, University of Milano-Bicocca, Milan, Italy}
\affil[2]{Department of Physics, University of Helsinki, Helsinki, Finland}
\date{}

\begin{document}
\maketitle

\begin{abstract}
$\beta$-\ce{Ga2O3} grows as three-dimensional islands on mismatched c-plane sapphire, often on a thin $\alpha$-\ce{Ga2O3} wetting layer, and eventually forms a continuous film by island coalescence. Experiments reveal the $\beta$ film to be relaxed. Surprisingly no dislocations are observed, suggesting an unusual plastic relaxation path. We reveal the key mechanisms underlying this process by combining density functional theory, continuum nucleation theory, and molecular dynamics simulations. Plastic relaxation occurs through rearrangement of the Ga atoms beneath the oxygen plane shared by both the wetting layer and the (-201) $\beta$-\ce{Ga2O3} structures. A local $\alpha$ to $\beta$-relaxed phase transition is thus realized within a single plane, allowing full relaxation of the $\beta$-\ce{Ga2O3} island. 
\end{abstract}

%\linenumbers
Gallium oxide (\ce{Ga2O3}) is an ultra-wide-bandgap semiconductor of increasing relevance for power electronics and optoelectronic applications~\cite{rev1}~\cite{rev2}~\cite{rev3}. Beyond its thermodynamically stable monoclinic $\beta$ phase, \ce{Ga2O3} exhibits multiple metastable polymorphs -- including $\alpha$, $\kappa$ and $\gamma$ -- all of which have been observed in heteroepitaxial growth on c-plane sapphire ($\alpha$-\ce{Al2O3}), sometimes coexisting within the same sample depending on growth conditions~\cite{doi:10.1021/acsami.5c13401}~\cite{Schewski_2015}. This behavior is promoted by a structural similarity of the oxygen sublattice in all \ce{Ga2O3} polymorphs and the substrate.

Under kinetically constrained regimes -- low substrate temperature and/or high deposition rate -- $\alpha$-\ce{Ga2O3} grows coherently in a layer-by-layer fashion, under a $\sim$4.5\% compressive misfit strain~\cite{ila1}, until plastic relaxation occurs through misfit dislocations~\cite{kaneko2012evaluation}~\cite{10.1063/5.0069554}~\cite{Takane_2021}~\cite{OSHIMA2021126387}. This behavior reflects both the structural correspondence of $\alpha$-\ce{Ga2O3} with the corundum substrate and the fact that, among the \ce{Ga2O3} polymorphs, $\alpha$ phase presents the lowest elastic energy under fully coherent conditions, as shown in Ref.~\cite{ila1} through calculations of volume, surface and interface energy including the effect of epitaxial strain. $\beta$-\ce{Ga2O3}, by contrast, is energetically disfavored because of its distinct crystal structure, exhibiting a larger and highly anisotropic lattice mismatch with $\alpha$ sapphire -- approximately 3.0\% and 9.5\% compressive strain along the two in-plane directions~\cite{ila1}. 
Accordingly, the direct epitaxial growth of the $\beta$ phase on sapphire is expected to be thermodynamically unfavorable. Nevertheless, it is experimentally observed as the dominant phase under intermediate-to-high deposition temperatures and/or low deposition rates, i.e. under conditions of high surface mobility~\cite{bosi2021thermodynamic}.
$\beta$-\ce{Ga2O3} nucleates and grows through 3D island formation~\cite{doi:10.1021/acsami.5c13401}~\cite{https://doi.org/10.1002/pssa.202000457}. In the early stages, growth proceeds through independent single-domain islands, rotated by 120°~\cite{doi:10.1021/acsami.8b17731}~\cite{nakagomi2012crystal}~\cite{wei2019beta} with respect to one another. A thin $\alpha$-\ce{Ga2O3} wetting layer is often observed at the film-substrate interface beneath the $\beta$ islands, independently of the growth technique~\cite{10.1063/1.4998804}~\cite{Schewski_2015}~\cite{feng2021investigation}. This three-dimensional morphology provides elastic relaxation through free surfaces, thereby making $\beta$ nucleation thermodynamically competitive despite its higher elastic energy. 
As the $\beta$ islands grow, however, the elastic energy accumulated increases and before islands merge into a continuous film, a different path of relaxation has to be activated. 
This additional relaxation is typically provided by misfit dislocations, which in conventional heteroepitaxy form in islands beyond a critical size, before coalescence~\cite{gatti2008modeling}. Surprisingly, this is not the case: experiments show that $\beta$ islands end up merging into a largely relaxed film in absence of dislocation formation~\cite{Schewski_2015}~\cite{doi:10.1021/acsami.5c13401}~\cite{10.1063/1.4998804}.

These observations place \ce{Ga2O3} heteroepitaxy outside the conventional framework of lattice-mismatched systems, where strain is accommodated within a single phase through well-established growth modes such as Frank-van der Merwe, Volmer-Weber, or Stranski--Krastanov. In \ce{Ga2O3}/sapphire, by contrast, the transition from $\alpha$ wetting layer to $\beta$ islands couples morphological change with polymorphic transformation, making the mechanism of strain accommodation during growth an intriguing open question.

In this work, we show that beyond elastic relaxation, an effective strain-relief pathway becomes accessible under high atomic mobility conditions: an atomic rearrangement confined to a single cation plane at the $\beta$/$\alpha$ interface, in which Ga atoms collectively redistribute into a Ga layer with alternating $\alpha$- and $\beta$-like coordination that accommodates the lattice-parameter discontinuity while preserving the oxygen framework.

To address this, we employ a multiscale approach combining Density Functional Theory (DFT), continuum nucleation theory and large-scale molecular dynamics simulations based on a machine-learned interatomic potential~\cite{gappot}. We provide a thermodynamic rationale for the formation of the $\alpha$ wetting layer and for the subsequent nucleation of $\beta$ islands, and we identify a novel strain-relief mechanism operating at the $\beta$/$\alpha$ interface. This mechanism -- fundamentally distinct from dislocation-mediated relaxation -- explains why a relaxed $\beta$-\ce{Ga2O3} film is observed on top of a strained wetting layer even beyond growth stages in which elastic surface relaxation is effective.

\section*{Results}

\begin{figure}[htbp]
    \centering
    \includegraphics[width=0.9\textwidth]{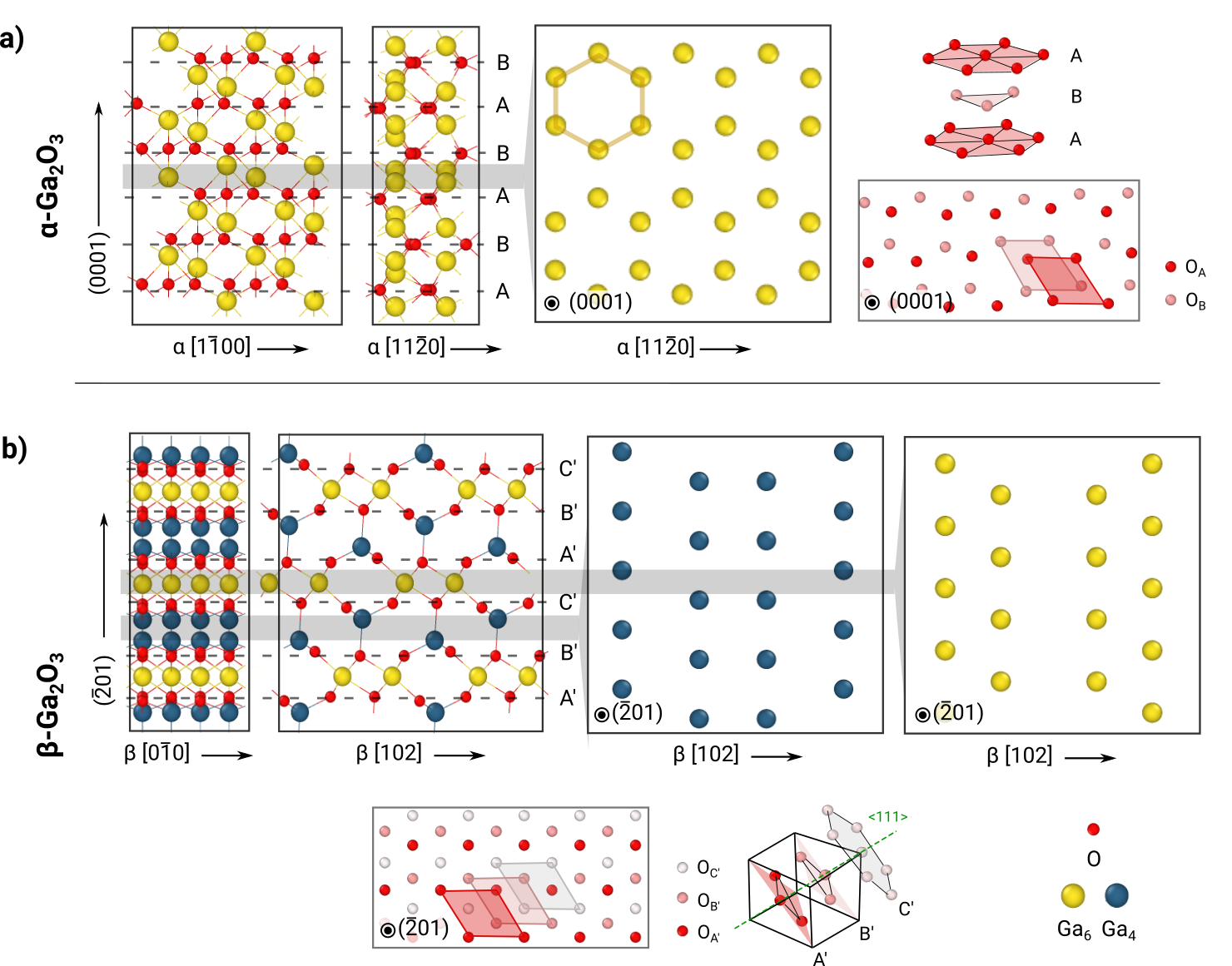}
    \caption{Oriented supercells of $\alpha$- (a) and $\beta$- (b) \ce{Ga2O3}
polymorphs. Left panels display front and side views of the cells and
top view of Ga layers. (a)-right and (b)-bottom panels show the HCP and
FCC oxygen sub-lattice of $\alpha$ and $\beta$ polymorphs respectively. Ga atoms are
represented by big spheres color-coded by their coordination numbers and
oxygen atoms are smaller red spheres.}
    \label{fig:struct}
\end{figure}

The following analysis is based on atomistic models of $\alpha$- and $\beta$-\ce{Ga2O3} in the experimentally observed growth orientations (Figure~\ref{fig:struct}a, b). The $\alpha$ phase is modeled with the (0001) growth front, whose oxygen sublattice consists of two hexagonal planes, A and B in HCP stacking, perpendicular to the growth direction; Ga atoms occupy octahedral sites between these layers (Figure~\ref{fig:struct}a).
$\beta$-\ce{Ga2O3} adopts the (-201) growth orientation, along which the oxygen framework follows an A\ensuremath{'} \textminus B\ensuremath{'} \textminus C\ensuremath{'} stacking sequence analogous to FCC stacking, with Ga atoms occupying alternating layers of octahedral and tetrahedral sites (Figure~\ref{fig:struct}b).

Coherent film/substrate interfaces are constructed by matching the in-plane oxygen sublattices while keeping the c-plane sapphire substrate fixed. For $\alpha$-\ce{Ga2O3} (0001) on $\alpha$-\ce{Al2O3} (0001), both compounds share the corundum structure, so interface construction is straightforward: the in-plane lattice parameters of $\alpha$-\ce{Ga2O3} are constrained to those of sapphire, which imposes an isotropic compressive strain of approximately 4.5\% along the $[1\bar{1}00]$ and $[11\bar{2}0]$ directions~\cite{ila1}. For $\beta$-\ce{Ga2O3}, we adopt the experimentally observed epitaxial relationships $\alpha$-\ce{Al2O3} $[1\bar{1}00]$ $\parallel$ $\alpha$-\ce{Ga2O3} $[1\bar{1}00]$ $\parallel$ $\beta$-\ce{Ga2O3} $[010]$ ($x$) and $\alpha$-\ce{Al2O3} $[11\bar{2}0]$ $\parallel$ $\alpha$-\ce{Ga2O3} $[11\bar{2}0]$ $\parallel$ $\beta$-\ce{Ga2O3} $[102]$ ($y$), with the hexagonal directions in four-index notation~\cite{Schewski_2015}. The $\beta$ (-201) oxygen framework is matched to the already strained $\alpha$ wetting layer, using the lowest-energy interface termination identified in Ref.~\cite{ila2}. The resulting mismatch is strongly anisotropic, amounting to $\sim$3.0\% along $[102]$ and $\sim$9.5\% along $[010]$. Unlike the isotropic $\alpha$ case, this anisotropy plays a central role in the strain-accommodation mechanism discussed below.

\subsection*{Island nucleation by continuum modeling}

\begin{figure}[htbp]
    \centering
    \includegraphics[width=0.5\textwidth]{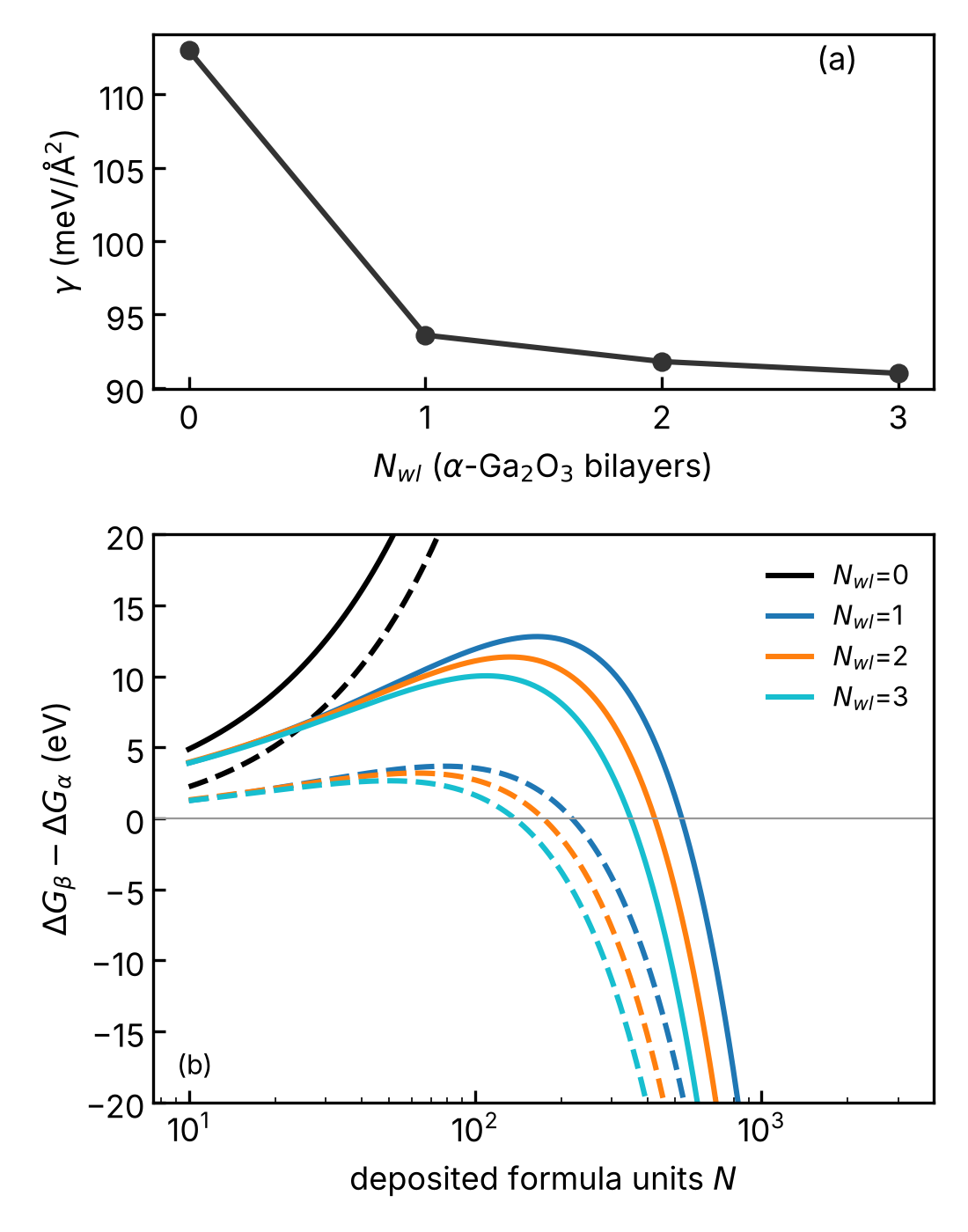}
    \caption{Surface energy $\gamma$ as a function of the number ($N$) of
deposited $\alpha$-\ce{Ga2O3} bi-layers over the $\alpha$-\ce{Al2O3} substrate (top image) and
the energy difference $\Delta G_{\beta-\alpha}$ as a function of the
number $N$ of deposited \ce{Ga2O3} formula units (bottom image, logarithmic scale): black curves refer to direct
deposition over $\alpha$-\ce{Al2O3} substrate, while blue, orange and light-blue
curves correspond to deposition over a surface already wetted by a
single, double and triple bi-layer of $\alpha$-\ce{Ga2O3}, respectively. Solid lines
correspond to the lower bound of the $\alpha$ step-edge energy (50~meV/Å), while
dashed lines correspond to its upper bound (100~meV/Å).}
    \label{fig:graph}
\end{figure}

We first address the thermodynamic driving force for the $\alpha$-\ce{Ga2O3} wetting of c-plane sapphire through DFT calculations. Periodic slabs of $\alpha$-\ce{Al2O3} (0001) were constructed with in-plane lattice parameters fixed to their bulk sapphire values. Successive unitary $\alpha$-\ce{Ga2O3} bi-layers were then added and the atomic positions were relaxed under epitaxial constraint. Bulk and interface contributions -- previously determined in Refs.~\cite{ila1}~\cite{ila2} -- were subtracted from the total slab energies to isolate the surface contribution as a function of the $\alpha$-\ce{Ga2O3} coverage (Figure~\ref{fig:graph}, top). The bare sapphire surface exhibits a high surface energy ($\sim$113 meV/Å²). Deposition of a single $\alpha$-\ce{Ga2O3} bi-layer already reduces this value, indicating a strong thermodynamic driving force for wetting.
This stabilization rapidly saturates: after approximately three bi-layers, the surface energy converges toward that of strained $\alpha$-\ce{Ga2O3} bulk, and further $\alpha$ growth provides no additional surface-energy gain.

The surface energy trend provides a clear thermodynamic rationale for the formation of a thin $\alpha$ phase interlayer, but alone does not explain the emergence of the $\beta$ phase. To clarify this, we analyze the nucleation competition between the two polymorphs, following the approach of Ref.~\cite{doi:10.1021/acsami.5c13401}. We evaluate the difference
$\Delta G_{\beta-\alpha}(N)=\Delta G_{\beta}(N)-\Delta G_{\alpha}(N)$ between the formation energies of a $\beta$ island and of an $\alpha$ platelet made of the same number $N$ of \ce{Ga2O3} formula units, on sapphire covered by $N_{wl}$ bi-layers of $\alpha$-\ce{Ga2O3}. Since both nuclei form from the same vapour, the chemical potential of the vapour enters both energies through the same term and cancels in their difference. The $\alpha$ phase is modeled as a fully strained two-dimensional platelet one bi-layer high, which extends the wetting layer from $N_{wl}$ to $N_{wl}+1$ bi-layers. Because of its small thickness, its surface contribution is given by the change of the surface energy of the wetting layer, $\gamma(N_{wl}+1)-\gamma(N_{wl})$ (Figure~\ref{fig:graph}a), while the cost of its edges is described by a step-edge energy $\lambda$, for which we consider the range 50--100~meV/Å~\cite{doi:10.1021/acsami.5c13401}. The free energy for the nucleation of $\alpha$-\ce{Ga2O3}, up to the term common to both nuclei, therefore reads:
\begin{equation}
\Delta G_{\alpha}(N) = \left(\mu_{\alpha}+\Delta\mu^{\varepsilon}_{\alpha}\right)N + \frac{v_{\alpha}}{h_{\alpha}}\left[\gamma(N_{wl}+1)-\gamma(N_{wl})\right]N + \lambda\, C_{p}\sqrt{\frac{v_{\alpha}}{h_{\alpha}}}\,\sqrt{N}
\label{eq:deltaGalfa}
\end{equation}

where $\mu_{\alpha}$ is the bulk energy per formula unit of $\alpha$-\ce{Ga2O3} relative to unstrained $\beta$-\ce{Ga2O3}, $\Delta\mu^{\varepsilon}_{\alpha}$ its elastic energy per formula unit under coherent strain, $v_{\alpha}$ the volume per formula unit, $h_{\alpha}$ the height of a single bi-layer, and $C_{p}$ the ratio between the perimeter of the platelet and the square root of its area (see the Methods section). The $\beta$ phase is instead modeled as a three-dimensional island shaped as a truncated cone with the experimentally suggested aspect ratio of $\sim$0.33~\cite{doi:10.1021/acsami.5c13401}, which allows a significant degree of elastic relaxation. In this case, the nucleation free energy of this polymorph reads:
\begin{equation}
\Delta G_{\beta}(N) = \Delta\mu^{\varepsilon}_{3D,\beta}\,N + \Gamma_{\beta}\, v_{\beta}^{2/3} N^{2/3}, \qquad
\Gamma_{\beta} = C_{top}\gamma_{top} + C_{lat}\gamma_{lat} + C_{bot}\left(\gamma_{int}-\gamma(N_{wl})\right)
\label{eq:deltaGbeta}
\end{equation}

where $\Delta\mu^{\varepsilon}_{3D,\beta}$ is the elastic energy retained by the island, 20.5\% of that of a coherent $\beta$ layer according to finite-element calculations (Figure~\ref{fig:strain}a), and $\Gamma_{\beta}$ is the sum of the energies of the top facet, of the lateral facets and of the base of the island, each weighted by its area divided by $V^{2/3}$ ($C_{top}$, $C_{lat}$, $C_{bot}$). The residual strain of the island is concentrated near its base, while its top is almost free of strain (Figure~\ref{fig:strain}a): the top facet is therefore described by the energy of the relaxed, unstrained (-201) surface~\cite{ila1}.

All input quantities entering the nucleation model -- bulk strain
energies, surface and interface energies, and elastic constants -- are
reported in the Supplementary Information.

The resulting $\Delta G_{\beta-\alpha}(N)$ curves are shown in
Figure~\ref{fig:graph}, bottom. On bare sapphire, $\Delta G_{\beta-\alpha}(N)$
remains positive for all sizes, confirming that the $\alpha$ phase is always thermodynamically preferred in the absence of a wetting layer. When nucleation is considered on top of an $\alpha$ wetting layer of 1--3 bi-layers, a crossover emerges: the $\alpha$ phase remains favorable at small sizes, where surface energy terms dominate, but the $\beta$ phase becomes advantageous above a critical size $N^*$, as the elastic energy gain from three-dimensional relaxation outweighs the surface energy cost. These results define a thermodynamic window in which coexistence of a thin $\alpha$ wetting layer and nucleating $\beta$ islands is energetically favorable.

Finite-element calculations show that $\beta$ islands significantly relax, particularly along the high-mismatch direction, confirming that three-dimensional morphology provides an efficient elastic-relief mechanism (Figure~\ref{fig:strain}a). 
 
This mechanism, however, is inherently transient. As the islands evolve into a continuous film upon further deposition elastic relaxation through the free surfaces progressively disappears.
Yet experiments still report the $\beta$ phase to be largely relaxed, in absence of misfit dislocations.
This discrepancy points to the existence of an additional strain-relief mechanism. To uncover it, we turn to large-scale molecular dynamics simulations using the machine-learning interatomic potential TABGAP~\cite{gappot}, previously shown to accurately capture extensive structural rearrangements and phase transformations among \ce{Ga2O3} polymorphs~\cite{nature}.

\subsection*{Island plastic relaxation by atomistic simulations}

\begin{figure}[htbp]
    \centering
    \includegraphics[width=0.95\textwidth]{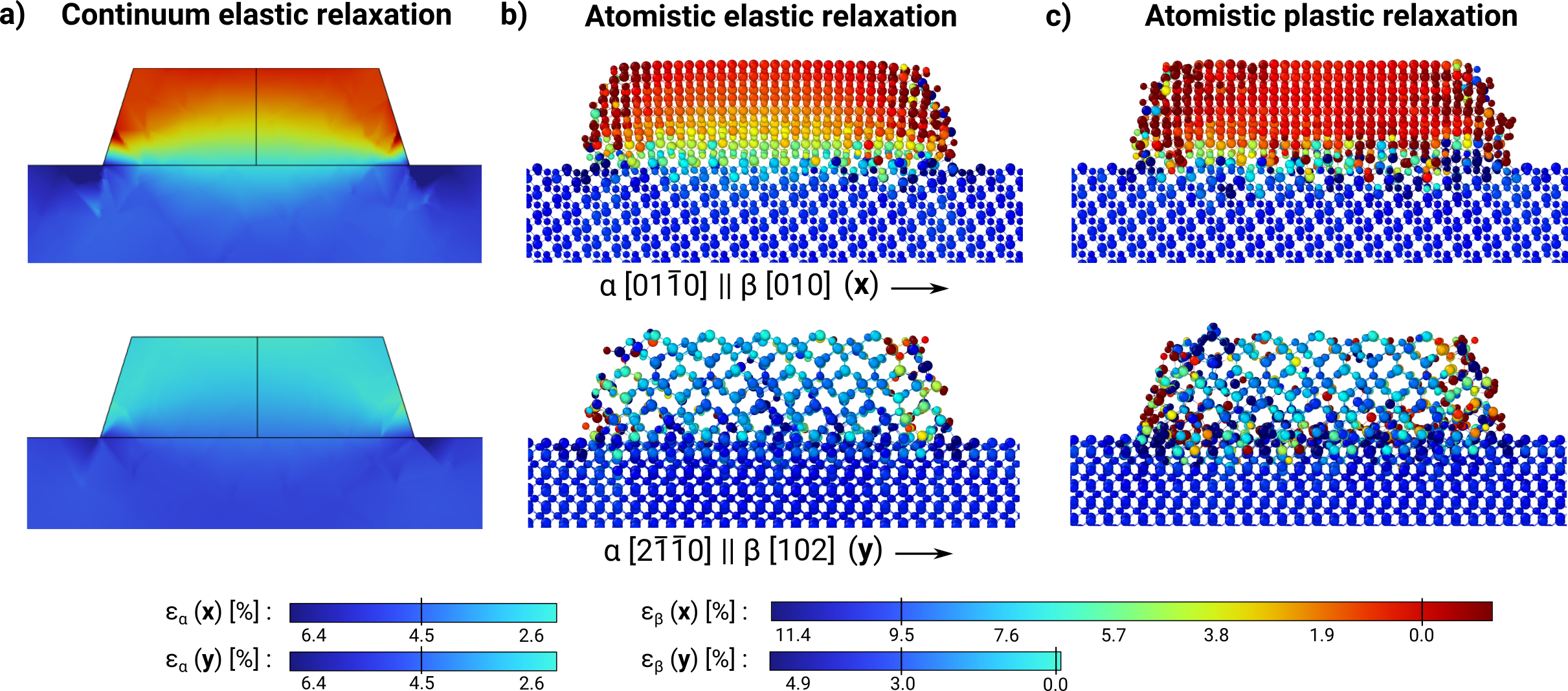}
    \caption{Strain color maps for the $\beta$-island/$\alpha$-wetting-layer
system after continuum elastic-, atomistic elastic- and atomistic
plastic-relaxation are reported by panels a), b) and c), respectively.}
    \label{fig:strain}
\end{figure}

A $\beta$-island/$\alpha$-wetting-layer heterostructure is constructed by placing a
truncated cone island of radius $\sim$2 nm -- with shape and aspect
ratio chosen to match the finite-element model discussed above -- on top of a fully
strained $\alpha$ wetting layer through a shared oxygen plane. The top surface
of the island adopts the well-established lowest-energy (-201)
termination~\cite{BERMUDEZ2006193}~\cite{doi:10.1021/acs.jpcc.0c00994}~\cite{10.1063/5.0019915}~\cite{ila1}. Energy minimization yields an elastically relaxed configuration in which strain is partially accommodated through free surfaces without structural rearrangements. This result closely matches the continuum prediction (Figure~\ref{fig:strain}a,b): the island core remains compressed while the outer regions relax toward the free surface.

We therefore extend our analysis to finite-temperature atomistic simulations. Molecular dynamics simulations of overcritical islands ran at 500 K reveal an additional strain-relief mechanism.

A collective atomic rearrangement occurs within the island, resulting in
a substantial reduction of the total potential energy relative to the
purely elastic configuration ($\sim$100 eV energy difference
for the case represented in Figure~\ref{fig:strain}b, c). This plastic relaxation allows the whole $\beta$ lattice to expand laterally towards its bulk relaxed state, while the underlying $\alpha$ wetting layer remains constrained to the substrate-imposed lattice parameters.
The resulting strain map (Figure~\ref{fig:strain}c) shows almost complete release of the
initial compressive strain. Unlike elastic relaxation, this plastic interfacial rearrangement permanently accommodates the lattice mismatch and therefore remains effective even after island coalescence, providing a strain-relief pathway that stabilizes the $\beta$ phase beyond the nucleation regime.

\section*{Discussion}

\begin{figure}[htbp]
    \centering
    \includegraphics[width=0.9\textwidth]{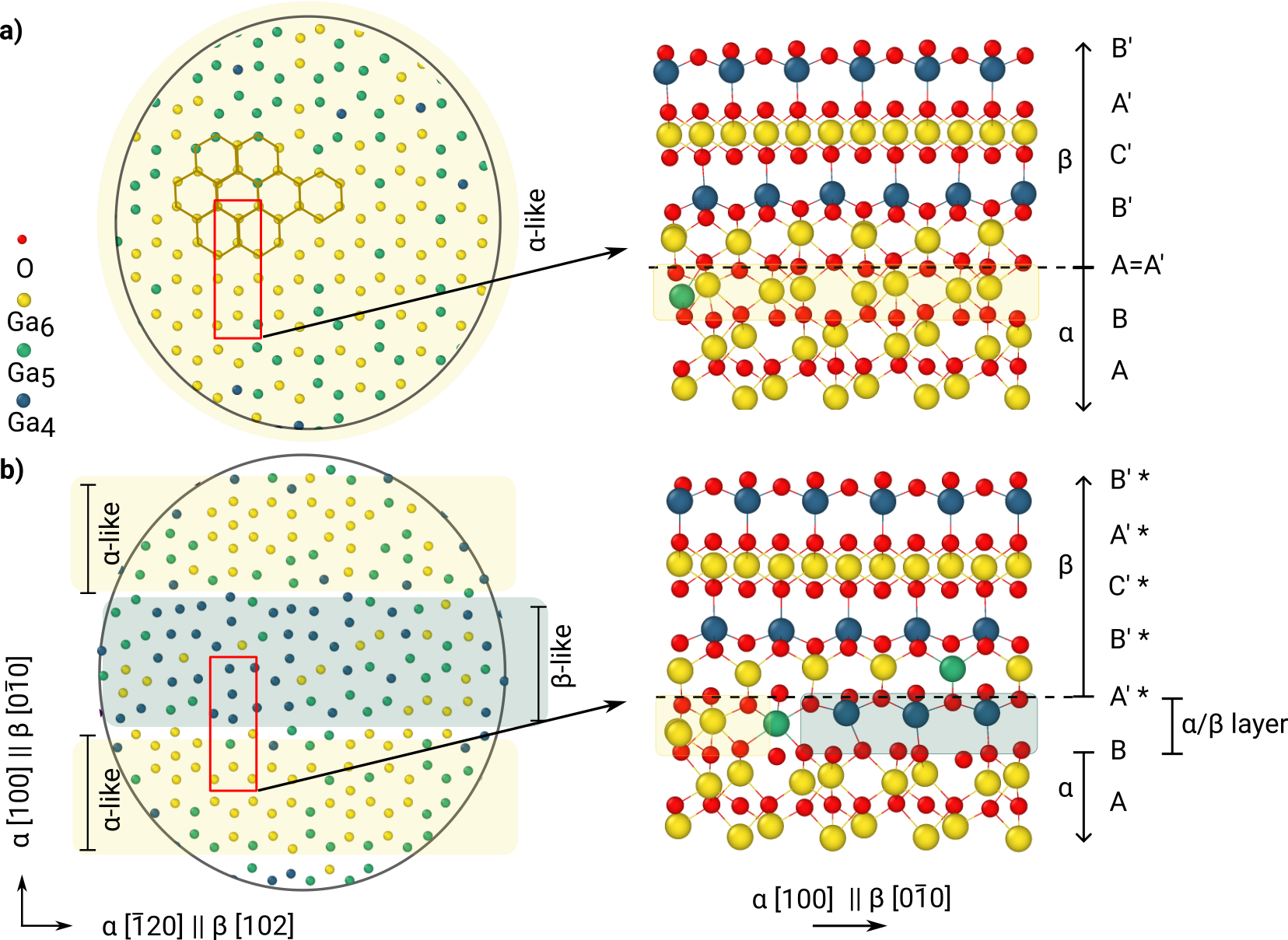}
    \caption{a) Front view of the atomistic configuration for the
adopted $\beta$ island/$\alpha$ wetting layer model, showing the coherent interface of oxygen
atoms (dashed line) between $\alpha$ wetting layer (bottom) and strained $\beta$
island (top). The black rectangle encloses the Ga plane involved in the
plastic relaxation-induced interfacial rearrangement. b) Top view of
the interfacial Ga layer after purely elastic relaxation; all Ga atoms
occupy $\alpha$-like positions. The red rectangle encloses a selected region
used for direct comparison with panel (c). Right: projection along the
$\alpha$-\ce{Ga2O3} $[1\bar{1}00]$ $\parallel$ $\beta$-\ce{Ga2O3} $[010]$ ($x$) direction of the
highlighted region, showing the coherent interface of oxygen atoms
(dashed line) between $\alpha$ wetting layer (bottom) and strained $\beta$ island
(top). c) Top view of the same Ga atoms after plastic relaxation.
Diffusion within the interfacial plane leads to spatial separation into
$\alpha$-like and $\beta$-like regions. The highlighted area captures a
representative boundary between these domains. Right: corresponding side
view showing the mixed Ga interlayer accommodating the lattice
discontinuity between relaxed $\beta$ (upper region) and strained $\alpha$ (lower
region). Oxygen stacking sequences are indicated; starred labels (A\ensuremath{'}*,
B\ensuremath{'}*, C\ensuremath{'}*) denote the relaxed $\beta$ sublattice, which no longer shares
periodicity with the $\alpha$ stacking. Ga atoms are represented by big
spheres color-coded by their coordination numbers and oxygen atoms are the smaller red spheres.}
    \label{fig:Ga}
\end{figure}

\begin{figure}[htbp]
    \centering
    \includegraphics[width=0.99\textwidth]{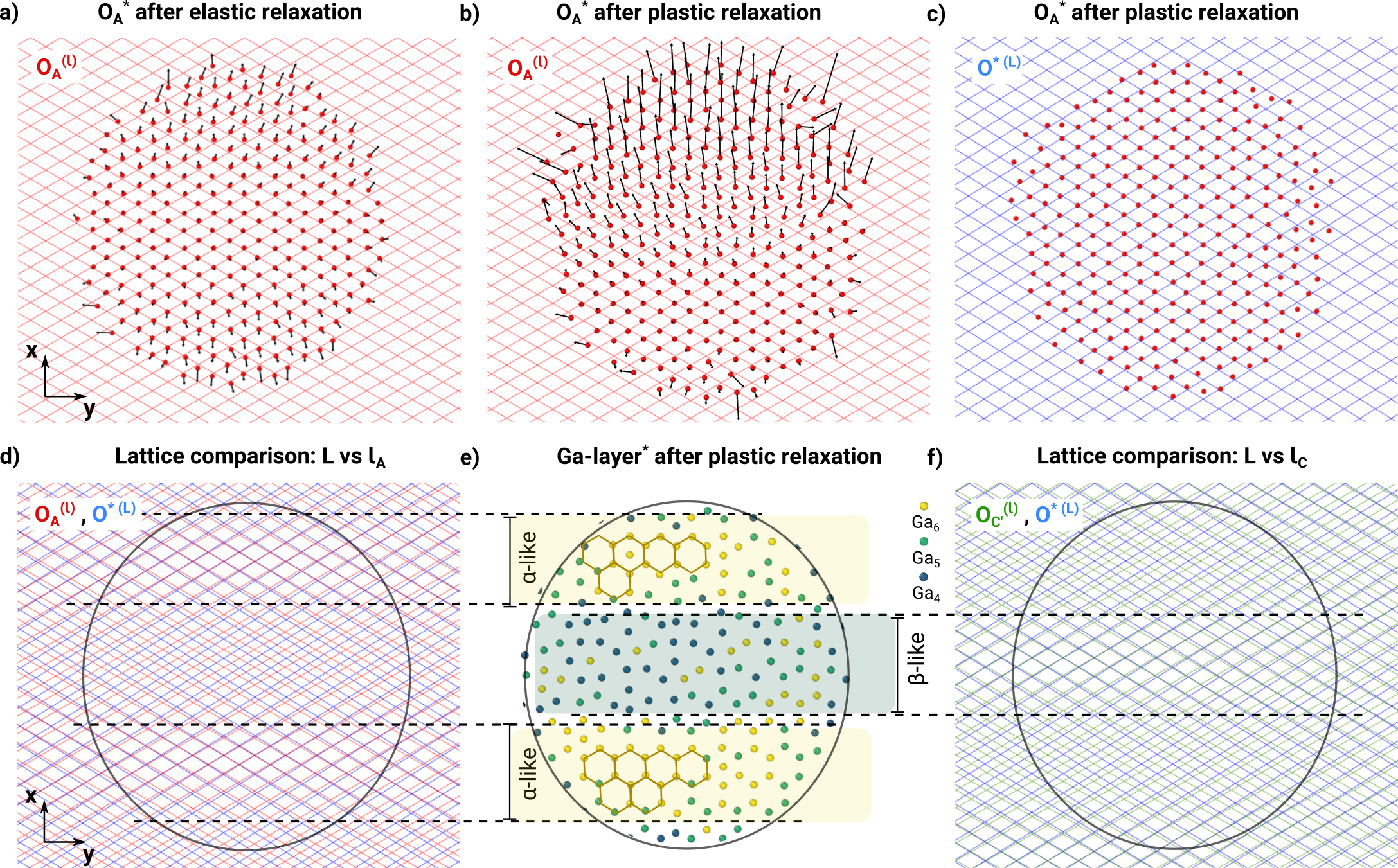}
    \caption{a) Oxygen atoms of the interface layer
(O\textsubscript{A}) after purely elastic relaxation, superimposed on
the initial strained $\alpha$ sublattice grid
O\textsuperscript{(l)}\textsubscript{A} (red lines); black arrows
indicate displacement magnitude and direction (amplified ×2 for
clarity). b) Same view after plastic relaxation. c) Oxygen atoms
post-plastic relaxation overlaid on the relaxed $\beta$ sublattice grid
O*\textsuperscript{($L$)} (blue), highlighting expansion toward the
relaxed lattice parameter. d) Schematic representation of the overlap
between initial $\alpha$ (red) and relaxed $\beta$ (blue) oxygen sublattices within
the island footprint (black circle), showing regions of local
coincidence along two bands. e) Interfacial Ga ``buffer'' layer beneath
the oxygen interface O*, showing alternating $\alpha$- and $\beta$-like Ga coordination which
accommodates the lattice mismatch between $\alpha$ wetting layer and relaxed $\beta$
island. Dashed lines guide the eye to panels (d) and (f). f) Schematic
of overlap between shifted $\alpha$ sublattice
O\textsuperscript{(l)}\textsubscript{C} (green) and relaxed $\beta$ grid
(blue), indicating regions where Ga adopts tetrahedral $\beta$-like
coordination. Panels (d--f) illustrate the periodic modulation of local
oxygen lattice matching and corresponding Ga coordination across the
interface.}
    \label{fig:Ox}
\end{figure}

To understand the microscopic origin of this plastic relaxation, we
analyze the atomic rearrangements occurring at the $\beta$/$\alpha$ interface. In the
initial coherent configuration, the $\beta$ island and the $\alpha$ wetting layer
share a common oxygen plane, whose in-plane lattice parameter is
constrained to the substrate value $l$. In this state the first oxygen
layer of the $\beta$ stacking (A \ensuremath{'}) coincides with the A layer of the
underlying $\alpha$ lattice (dashed line of Figure~\ref{fig:Ga}a), ensuring epitaxial
registry between the two oxygen frameworks. After purely elastic
relaxation, this registry is preserved: while the island tends to expand
to relieve the compressive strain, the epitaxial constraint limits
atomic rearrangements. This is consistently captured by both Figure~\ref{fig:Ga}b and
Figure~\ref{fig:Ox}a -- the interfacial Ga layer shows all atoms occupying $\alpha$-like
positions, and the interfacial oxygen atoms exhibit small displacement
vectors that largely preserve the A \ensuremath{'}$\sim$ A registry, predominantly aligned
along the high-mismatch direction ($\sim$9.5\% compressive
strain), reflecting the anisotropic elastic energy accumulated in the
coherently strained island.

Upon plastic relaxation, the $\beta$ island expands toward its bulk lattice
parameters and its entire oxygen framework --- including the interfacial
plane --- relaxes to the sublattice parameter L, becoming
O*\textsuperscript{($L$)} (Figure~\ref{fig:Ox}b, c). The atomic displacements are now
substantially larger and no longer centered around the coherent lattice
positions, so that the interfacial oxygen lattice loses commensurability
with the $\alpha$ wetting layer beneath it. The oxygen stacking above the
interface thus follows the relaxed $\beta$ sequence A\ensuremath{'}*, B\ensuremath{'}*, C\ensuremath{'}*, while the $\alpha$ wetting layer below retains the A, B stacking at parameter $l$
(Figure~\ref{fig:Ga}c, left). This lattice-parameter discontinuity is the driving
force for the rearrangement of the Ga layer immediately beneath the
interfacial oxygen plane (Figure~\ref{fig:Ga}a, black rectangle). Since the relaxed
$\beta$ oxygen sublattice (parameter $L$) is expanded relative to its initial
fully strained configuration (parameter $l$), the two grids periodically
come into and out of registry along the high-mismatch direction. Where
they coincide, the interfacial oxygen plane sits above
O\textsuperscript{($l$)}\textsubscript{A} sites --- the same HCP-like
configuration as in the coherent interface --- and the Ga atoms beneath
adopt $\alpha$-like octahedral coordination (Figure~\ref{fig:Ox}d). Where registry is lost,
the interfacial oxygen plane instead finds coincidence with
O\textsuperscript{($l$)}\textsubscript{C} sites, imposing a locally
FCC-like environment that drives the underlying Ga atoms into $\beta$-like
tetrahedral coordination (Figure~\ref{fig:Ox}f). The result is alternating $\alpha$- and $\beta$-like Ga layer
(Figure~\ref{fig:Ga}c, Figure~\ref{fig:Ox}e) that mediates the strain discontinuity within a single
atomic plane while preserving long-range oxygen crystallinity.
The atomic-scale confinement of this rearrangement, together with the preservation of the oxygen framework, makes its direct identification by conventional cross-sectional TEM challenging and may account for the lack of previous reports of similar strain-accommodation mechanisms.

It is worth noting that this rearrangement mechanism is fully
compatible with the experimentally observed formation of rotational $\beta$
domains separated by $\sim$120° on the hexagonal $\alpha$ surface.
Prior to plastic relaxation, $\beta$ islands nucleate coherently on the $\alpha$
wetting layer with orientations related by the three-fold symmetry of
the underlying hexagonal oxygen sublattice. The plastic rearrangement
does not alter these orientational relationships: while the interfacial
oxygen layer loses registry with the strained $\alpha$ sublattice, the
hexagonal symmetry of the oxygen planes is preserved throughout, and the
relaxed $\beta$ lattice retains the same in-plane orientation as the coherent
island from which it evolved. The rotational domain structure
established during nucleation is therefore unaffected by the interfacial
rearrangement, remaining consistent with experimental observations.

This interfacial plasticity explains how $\beta$ can remain nearly fully
relaxed on top of a strained $\alpha$ wetting layer even in the absence of
misfit dislocations, in agreement with experimental observations. Unlike
conventional heteroepitaxy -- where lattice mismatch is accommodated
through dislocation networks within the same crystalline phase -- the
present oxide system relieves strain via a localized cation
redistribution confined to a single crystallographic plane, while preserving the crystalline continuity of the oxygen framework across the
interface. This behavior reflects a key structural feature of
polymorphic oxides: the rigidity of the close-packed anion sublattice,
combined with the flexibility of cation coordination across different
stacking environments, enables strain accommodation through
phase-selective cation rearrangements without disrupting the oxygen
framework. Strain-driven interfacial rearrangements of this kind may
therefore represent a general, underexplored relaxation channel in
complex oxide heteroepitaxy.

\section*{Methods}

First-principles calculations were performed using the Vienna Ab initio
Simulation Package (VASP)~\cite{Kresse1996a}~\cite{Kresse1996b} within density functional theory. The
Perdew--Burke--Ernzerhof revised for solids (PBEsol)~\cite{Perdew2008} functional within
the generalized gradient approximation (GGA) was employed, and
projector-augmented wave (PAW)~\cite{Kresse1999} potentials were used for Ga, O, and Al,
explicitly including Ga d electrons. A plane-wave cutoff of 500\,eV and
a 6×6×1 Monkhorst--Pack k-point mesh~\cite{Monkhorst1976} were used for all slab calculations.
Periodic slab models were constructed with the x--y in-plane dimensions
fixed to the conventional $\alpha$-\ce{Al2O3} lattice, oriented so that the surface
perpendicular to z corresponds to the (0001) termination. Stoichiometric
slabs consisted of seven sapphire bi-layers with N additional $\alpha$-\ce{Ga2O3}
bi-layers on top, with $N = 1-10$, where we chose nine bi-layers as the
smallest thickness granting independent surfaces, based on a previous
work of ours~\cite{ila1}. We separated each slab from its
periodic images by inserting a vacuum region of at least 15\,Å along the
z direction. Atoms in the bottom seven bi-layers of sapphire were frozen
to preserve substrate structure, while the remaining sapphire layers
near the interface and all $\alpha$-\ce{Ga2O3} atoms were fully relaxed. In order to estimate the thickness-dependent surface energies of the epitaxial film,
we calculated the energy differences relative to the thickest $\alpha$-\ce{Ga2O3} slab by removing a stoichiometric $\alpha$-\ce{Ga2O3} layer at a time, according to the following relationship:

\begin{equation}
\Delta E(N^{*}) = E_{slab}(N^{max}) - E_{slab}(N^{*}) - \Delta N \cdot E_{bi-layer}
\label{eq:deltaE}
\end{equation}

where $\Delta N = N^{max} - N^{*}$, and E\textsubscript{bi-layer} is the energy of a bulk
$\alpha$-\ce{Ga2O3} (stoichiometric) bi-layer. The interface energy term is zero
based on our previous interface-energy analysis~\cite{ila2}, hence we assume only the surface energy
contributes to $\Delta E$: this allows us to obtain the
surface energy as $\gamma(N)\ = \gamma_{N^{max}} + \Delta E(N)/A$, with $A$ the in-plane area of the slab, which
is reported in Figure~\ref{fig:graph}a. This provides a consistent and systematic
measure of the surface stabilization associated with successive bi-layer
deposition on $\alpha$-\ce{Al2O3}.

In the nucleation model, the perimeter factor $C_{p}$ of the $\alpha$ platelet depends on its in-plane shape, which is not known. For compact shapes it varies between $2\sqrt{\pi} \approx 3.54$ for a disc, the isotropic limit with the shortest perimeter for a given area, and 4 for a square, a representative faceted shape with straight edges and corners. The average of these two limits, $C_{p} = 3.77$, was used.

The elastic relaxation mechanism of three-dimensional $\beta$-\ce{Ga2O3} islands
placed on a strained $\alpha$-\ce{Ga2O3} wetting layer was investigated through
finite-element method simulations, performed using COMSOL
Multiphysics suite~\cite{comsol64}. The $\beta$ islands were modeled as truncated cones of top radius 22~Å, bottom radius 27.5~Å and height 18~Å placed on a planar $\alpha$-\ce{Ga2O3} substrate.
Elastic constants for $\alpha$- and $\beta$-\ce{Ga2O3} were taken from DFT calculations,
with a methodology described at Ref.~\cite{doi:10.1021/acsami.5c13401}: the values we
calculated are reported in the Supplementary Information. We set the eigenstrain of the two
materials to the values of the lattice misfit of $\alpha$-\ce{Ga2O3} (4.5\%) and $\beta$-\ce{Ga2O3} (3.0\% along $[102]$ and 9.5\% along $[010]$)
with c-sapphire, computed with PBEsol~\cite{ila1}.

Atomistic molecular dynamics simulations were performed using the LAMMPS
package~\cite{Thompson2022} with the machine-learning interatomic potential TABGAP for \ce{Ga2O3}
polymorphs \cite{gappot}. The simulation cell contained $\sim$88,000 atoms and spanned
approximately 180\,Å × 180\,Å in the x--y plane. $\beta$ islands were placed
on a strained $\alpha$-\ce{Ga2O3} wetting layer via a shared oxygen interface,
reproducing the experimentally observed epitaxial orientation. The in-plane strains
of the PBEsol calculations (4.5\% for $\alpha$; 3.0\% along $[102]$ and 9.5\% along $[010]$ for $\beta$) were applied to the equilibrium lattices of the potential. Only the upper six
bi-layers of the $\alpha$ wetting layer were left free to move during our
simulations; atoms below this region were fixed to mimic substrate
epitaxial constraints. Periodic boundary conditions were applied
in-plane (x and y direction) only.

The initial configuration was first relaxed to a local minimum of the
potential energy surface, corresponding to elastic relaxation, in which
the system accommodates strain without significant structural
rearrangement. Subsequently, NVT molecular dynamics simulations, with a Nos\'e--Hoover thermostat (damping time 80\,fs), were carried
out for a total of 600\,ps with a timestep of 1\,fs. The simulation
protocol consisted of three stages: (i) a gradual heating ramp from 0\,K
to 500\,K (200 ps), (ii) thermalization at 500\,K for 200\,ps, and (iii)
a gradual quench to 0\,K (200 ps), followed by a final energy
minimization. The gradual heating ensures that the system is brought to
the target temperature without imparting a sudden kinetic energy
``shock,'' which could cause the system to skip intermediate states or
access unphysical configurations. The thermalization stage allows the
system to explore the configurational space around 500\,K and relax
vibrational degrees of freedom. The slow quenching aims at avoiding the trapping of the 
system in high-energy local minima that could result from directly
minimizing a 500\,K snapshot. This procedure enables the system to reach
a deeper, more physically meaningful minimum, corresponding to plastic
relaxation and allowing interface rearrangement between the $\beta$ island
and the $\alpha$-\ce{Ga2O3} wetting layer.

Atomic trajectories were analyzed using the software OVITO~\cite{Stukowski_2010} to
extract structural properties of the system. In particular, the
coordination number of Ga atoms was counted with a cutoff radius of
2.5\,Å, and local strain distribution of the relaxed configurations was
evaluated to characterize deformation in the $\alpha$-\ce{Ga2O3} wetting layer and
the $\beta$ island.

\printbibliography
\end{document}